# Pairing and Phase Separation in a Polarized Fermi Gas

Guthrie B. Partridge, Wenhui Li, Ramsey I. Kamar, Yean-an Liao, Randall G. Hulet

Department of Physics and Astronomy and Rice Quantum Institute

Rice University

Houston, Texas 77251

We report the observation of pairing in a gas of atomic fermions with unequal numbers of two components. Beyond a critical polarization, the gas separates into a phase that is consistent with a superfluid paired core surrounded by a shell of normal unpaired fermions. The critical polarization diminishes with decreasing attractive interaction. For near zero polarization, we measure the parameter $\beta = -0.54 \pm 0.05$ describing the universal energy of a strongly interacting paired Fermi gas, and find good agreement with recent theory. These results are relevant to predictions of exotic new phases of quark matter and of strongly magnetized superconductors.

Fermion pairing is the essential ingredient in the Bardeen, Cooper, and Schrieffer (BCS) theory of superconductivity. In conventional superconductors, the chemical potentials of the two spin-states are equal. There has been great interest, however, in the consequences of mismatched chemical potentials which may arise in several important situations, including, for example, magnetized superconductors (*1-3*), or cold dense quark matter at the core of neutron stars (*4*). A chemical potential imbalance may be produced by several mechanisms, including magnetization in the case of superconductors, mass asymmetry, or unequal numbers. Pairing is qualitatively altered by the Fermi energy mismatch, and there has been considerable speculation regarding the nature and relative stability of various proposed exotic phases. In the Fulde-Ferrel-Larkin-Ovchinnikov (FFLO) phase (*2, 3*), pairs possess a non-zero center-of-mass momentum which breaks translational invariance, while the Sarma (*1*), or breached pair phase (*5*), is speculated to have gapless excitations. A mixed phase has also been proposed (*6-8*), in which regions of a paired BCS superfluid are surrounded by an unpaired normal phase. Little is known experimentally, however, because of the difficulty in creating magnetized superconductors. Initial evidence for an FFLO phase in a heavy-fermion superconductor has only recently been reported (*9, 10*). Opportunities for experimental investigation of exotic pairing states have expanded dramatically, however, with the recent realization of the BEC-BCS crossover in a two spin-state mixture of ultracold atomic gases. Recent experiments have demonstrated both superfluidity (*11-13*) and pairing (*14-17*) in atomic Fermi gases. We report the observation of pairing in a polarized gas of $^6$Li atoms. Above an interaction-dependent critical polarization, we observe a phase separation that is consistent with a uniformly paired superfluid core surrounded by an unpaired shell of the excess spin state. Below the critical polarization, the spatial size of the gas is in agreement with expectations for a universal, strongly-interacting paired Fermi gas.

Our methods for producing a degenerate gas of fermionic $^6$Li atoms (*18, 19*) and the realization of the BEC-BCS crossover at a Feshbach resonance (*17*) have been described previously (*20*). An incoherent spin mixture of the $F = ½$, $m_F = ½$ (state $|1\rangle$) and the $F = ½$, $m_F = -½$ (state $|2\rangle$) sublevels (where $F$ is the total spin quantum number and $m_F$ is its projection) is created by radio frequency (rf) sweeps, where the relative number of the two states can be controlled by the rf power (*20*). The spin mixture is created at a magnetic field of 754 G, which is within the broad Feshbach resonance located near 834 G (*21, 22*). The spin mixture is evaporatively cooled by reducing the depth of the optical trap that confines it, and the magnetic field is ramped adiabatically to a desired field within the crossover region. States $|1\rangle$ and $|2\rangle$ are sequentially and independently imaged in the trap by absorption (*20*). Analysis of these images provides measurement of $N_i$ and polarization $P = (N_1 - N_2) / (N_1 + N_2)$, where $N_i$ is the number of atoms in state $|i\rangle$. We express the Fermi temperature $T_F$ in terms of the majority spin state, state $|1\rangle$, as $k_B T_F = \hbar \bar{\omega} (6N_1)^{1/3}$, where $\bar{\omega} = 2\pi (v_r^2 v_z)^{1/3}$ is the mean harmonic frequency of the cylindrically symmetric confining potential with radial and axial frequencies, $v_r$ and $v_z$, respectively. For $P \approx 0$, we find that $N_1 \approx N_2 \approx 10^5$, giving $T_F \approx 400$ nK for our trap frequencies. Due to decreasing evaporation efficiency with increasing polarization, there is a correlation between $P$ and total atom number (Fig. S1).

For fields on the low-field (BEC) side of resonance, real two-body bound states exist and molecules are readily formed by three-body recombination. For the case of $P = 0$, a molecular Bose-Einstein condensate (MBEC) is observed to form with no detectable thermal molecules (*17*). Based on an estimated MBEC condensate fraction of >90%, we place an upper limit on the temperature $T < 0.1$ $T_F$ at a field of 754 G (*17*). However, the gas is expected to be cooled further during the adiabatic ramp for final fields greater than 754 G (*17*). Using similar experimental methods, we previously measured the order parameter of the gas in the BCS regime and found good agreement with T = 0 BCS theory (*17*), indicating that the gas was well below the critical temperature for pairing.

Figure 1 shows images of states |1⟩ and |2⟩ at a field of 830 G, for relative numbers corresponding to $P = 0.14$. The strength of the two-body interactions is characterized by the dimensionless parameter, $k_F a$, where $k_F$ is the Fermi wavevector and $a$ is the *s*-wave scattering length. For a field of 830 G, $k_F a > 10$, corresponding to a unitarity limited interaction. As discussed below, we contend that the gas has separated into a uniformly paired, unpolarized inner core surrounded by a shell of the excess, unpaired state |1⟩ atoms. In this case, the distribution of the difference, |1⟩ - |2⟩, also shown in Fig. 1, represents the location of these unpaired state |1⟩ atoms.

Figure 2 shows axial profiles of a sequence of images corresponding to increasing values of *P*, again for 830 G. These axial profiles are the result of integrating the column density over the remaining radial coordinate. They are insensitive to the effect of finite imaging resolution in the radial dimension, as well as to probe-induced radial heating of the second image in the sequence (*20*). On the left of the figure are distributions for both states |1⟩ and |2⟩, while the right side shows the corresponding difference distributions. Also shown in Fig. 2 are fits to a non-interacting $T = 0$ integrated Thomas-Fermi (T-F) distribution for fermions, $A\left(1 - \frac{z^2}{R^2}\right)^{\frac{5}{2}}$, where *A* and *R* are adjustable fitting parameters, and z is the axial position. Although the distributions are expected to differ somewhat from that of a non-interacting Fermi gas, we find that the fits are qualitatively good and provide a useful measure of the spatial size of the distributions. For $P = 0$ (Fig. 2A), the two spin components have identical distributions. We previously found that the gas was paired under the same conditions (*17*). As *P* increases (Fig. 2B), the peak height and width of the state $|2\rangle$ distributions initially diminish with respect to state $|1\rangle$, but their shapes are not fundamentally altered. When the polarization is increased beyond a critical value, however, the shapes of the two clouds become qualitatively different (Fig. 2C): The inner core, reflected by the distribution of the $|2\rangle$ atoms, is squeezed and becomes taller and narrower. This narrowing is noticeable in the wings of the state $|2\rangle$ distribution in comparison with the T-F fit. The squeezing of the state $|2\rangle$ distribution is accompanied by the excess, unpaired state $|1\rangle$ atoms being expelled from the center of the trap. These unpaired atoms form a shell that surrounds the inner core. As *P* approaches 1 (Fig. 2D), the contrast in the center hole in the difference distribution decreases because of the contribution to the axial density of unpaired atoms in the shell surrounding the core. The observation of difference distributions with a center hole and two peaks on either side is consistent with phase separation. Although more exotic redistributions of atoms cannot be ruled out, a separation between a uniformly paired phase and the excess unpaired atoms is the simplest explanation, and is consistent with theoretical predictions (*6-8*).

To gain a more quantitative understanding of the phase separation as a function of *P* we plot the ratio $R/R_{TF}$ vs. *P*, where $R_{TF} = \left(\frac{2k_B T_F}{m\omega_z^2}\right)^{\frac{1}{2}}$ is the axial Thomas-Fermi radius for non-interacting fermions (*23*), and where *m* is the atomic mass, $\omega_z = 2\pi\upsilon_z$, and $T_F$ is calculated for each state from the measured numbers $N_1$ and $N_2$. Figure 3 shows the results for all of the 830 G data. It is seen that at a critical polarization, $P_c = 0.09 \pm 0.025$, $R/R_{TF}$ for states $|1\rangle$ and $|2\rangle$ diverge in opposite directions from their value at small *P*. $R/R_{TF}$ for state $|2\rangle$, which corresponds to the distribution of the pairs, decreases continuously to ~0.4 for the maximum attained polarization of $P \sim 0.86$. For state $|1\rangle$, $R/R_{TF}$ jumps from its initial value to near unity at the critical polarization. Because $P = 1$ corresponds to a non-interacting gas, one expects $R/R_{TF}$ to approach unity in this limit.

In the case of $P \approx 0$, the observation that the axial extent of the paired cloud is smaller than that of a non-interacting Fermi gas can be explained by the universal energy of strongly interacting paired fermions at the unitarity limit, where $k_F|a| \gg 1$ (*24*). In this limit, the chemical potential of the gas is believed to have the universal form $E_F(1 + \beta)^{\frac{1}{2}}$, where $\beta$ is a universal many-body parameter which can be determined from $\beta = (R/R_{TF})^4 - 1$ (*25-27*). For $P$ near zero, we find that $R/R_{TF} = 0.825 \pm 0.02$, giving $\beta = -0.54 \pm 0.05$ (uncertainties discussed in Fig. 3 caption). This value is in excellent agreement with previous measurements (*24, 26, 28, 29*), but with significantly improved uncertainty. Our measurement is also consistent with $\beta = -0.58 \pm 0.01$ obtained from two Monte Carlo calculations (*8, 30, 31*) and with $\beta = -0.545$ from a calculation reported in (*27*). Not surprisingly, the measurement is in disagreement with $\beta = -0.41$ obtained with BCS mean-field theory (*27*).

We believe that the data are consistent with a quantum phase transition from a homogenous paired superfluid state to a superfluid-normal phase separated state. For $P = 0$, the excellent agreement between the measured value of $\beta$ and theory, combined with our previous measurement of pair correlations in an unpolarized gas (*17*) is strong evidence that the gas is paired. Furthermore, superfluidity has been observed in the same system under similar conditions (*11-13*). The fact that the size of the gas, which is strongly dependent on the gas being paired, does not change appreciably for $0 < P < P_c$, suggests that it may remain paired in this regime, which is remarkable (*32*). For $P > P_c$, the excess unpaired atoms prefer to reside in a shell outside the inner core. Such a phase separation may be explained in the BEC regime (*33*) where the atoms and weakly-bound dimers are believed to have a large repulsive three-body interaction (*34*), however, application of this theory to the strongly-interacting regime would be incorrect because it also gives a large repulsive dimer-dimer interaction (*34*) that is inconsistent with a negative value of β. Therefore, we conclude that the phase separation is a consequence of the energy cost of accommodating unpaired atoms within the paired core (*6-8*). Vortices have also been used to explore superfluidity in $^6$Li with mismatched Fermi surfaces (*35*). Although hints of phase separation are reported in that work, a critical polarization was not observed.

We have also performed the experiment at 920 G, which is on the BCS side of the resonance where $k_F a$ = -1.1. We find a phase separation at this field as well. However, the value of $R/R_{TF}$ at $P \approx 0$ is larger, 0.92 ± 0.02, a consequence of smaller but still strong interactions, and the critical polarization for phase separation is considerably smaller, $P_c$ < 0.03 consistent with zero to within our experimental sensitivity. Observation of phase separation at small $P$ demonstrates the sensitivity of our determination of phase separation. In the BEC regime at a field of 754 G, where $k_F a$ = 0.6, we find that $P_c$ is somewhat larger than 0.10, but at this field probe-induced radial heating prevents an accurate determination (*20*). For samples prepared at higher temperature ($T \approx 0.7\ T_F$), no phase separation was observed.

We report the observation of a phase transition from a uniform superfluid to a phase that is consistent with a segregated superfluid/normal state when the polarization exceeds a critical value. This critical value diminishes going from the BEC to BCS regimes, as expected (*8*). In the BCS regime, very little Fermi energy mismatch is tolerated before phase separation occurs. The nature of the coexistence phase where $P < P_c$ is still unknown, so the existence of the FFLO and breached pair states are not excluded by these observations. Recent calculations suggest that a homogeneous gapless superfluid state may be preferred for small polarizations in the unitarity regime (*8*). These results help to clarify the long open question of how Fermi superfluids respond to mismatched Fermi surfaces.

**Supporting Online Material**
www.sciencemag.org
Materials and Methods
Fig. S1

**Captions**

**Fig 1.** *In situ* absorption images showing phase separation at a field of 830 G. A false-color scale is used to represent the column density. The trapping frequencies are $\upsilon_r = 350$ Hz and $\upsilon_z = 7.2$ Hz. These images correspond to $P = 0.14$. **(A)** Majority spin state, $|1\rangle$, with $N_1 = 8.6 \times 10^4$. **(B)** Minority spin state, $|2\rangle$, with $N_2 = 6.5 \times 10^4$. **(C)** Difference distribution, $|1\rangle - |2\rangle$, corresponding to the excess unpaired $|1\rangle$ atoms. These excess atoms reside in a shell surrounding an inner core of unpolarized pairs. We observe that the excess state $|1\rangle$ atoms preferentially reside at large z, while relatively few occupy the thin radial shell at small z. We speculate that this may be a consequence of the high aspect ratio trapping potential. **(A)** and **(B)** were obtained sequentially using probe laser beams of different frequency. Probe-induced radial heating of the second image in the sequence (state $|1\rangle$, in this case), caused by off-resonant excitation by the first probe, produces a slight reduction in peak height (*20*). As a result, the difference distribution is slightly negative at the center. The size of each image in the object plane is 1.41 mm horizontally and 0.12 mm vertically. The displayed aspect ratio has been rescaled for clarity.

**Fig. 2.** Axial density profiles at 830 G. For the curves on the left, the blue (red) data correspond to state $|1\rangle$ ($|2\rangle$), while the green curves on the right show the difference distributions, $|1\rangle - |2\rangle$. The axial density measurements are absolute and without separate normalization for the two states. The solid lines on the left curves are fits to a Thomas-Fermi distribution for fermions, where the fitted parameters are *A* and *R*. **(A)** $P = 0.01$, $N_1 = 6.4 \times 10^4$; **(B)** $P = 0.09$, $N_1 = 1.0 \times 10^5$; **(C)** $P = 0.14$, $N_1 = 8.6 \times 10^4$; **(D)** $P = 0.53$, $N_1 = 6.8 \times 10^4$. The state $|2\rangle$ distributions reflect

the distribution of pairs, while the difference distributions show the unpaired atoms. Phase separation is evident in (**C**) and (**D**). The profiles in (**C**) are derived from the images given in Fig. 1.

**Fig. 3.** $R/R_{TF}$ vs. $P$. The ratio of the measured axial radius to that of a non-interacting Thomas-Fermi distribution are shown as blue open circles for state $|1\rangle$, and red crosses for state $|2\rangle$. The data combine 92 independent shots. The dashed line corresponds to the estimated critical polarization, $P_c = 0.09$, for the phase transition from coexisting to separated phases. The images are of sufficient quality that the assignment of phase separation is ambiguous in only two of the shots represented in this figure. Our contention for a phase transition at $P_c$ is based on statistical evidence: none of the 31 shots deliberately prepared as $P = 0$ and only one with a measured $P < 0.07$ are phase separated, while all but two shots with $P > 0.11$ are. The width of this transition region is consistent with our statistical uncertainty in the measurement of $P$. Although fluctuations in absolute probe detuning lead to 15% uncertainty in total number, the difference in the two probe frequencies is precisely controlled, resulting in lower uncertainty in $P$. We estimate the uncertainty in a single measurement of $P$ to be 5%, which is the standard deviation of measurements of $P$ for distributions prepared as $P = 0$. Also from these distributions, we find no significant systematic shift in detection of relative number. The uncertainty in the ratio $R/R_{TF}$ is estimated to be 2.5%, due mainly to the uncertainty in measuring $\upsilon_z$ (*20*). The uncertainty in $R/R_{TF}$ for state $|2\rangle$ grows with increasing $P$ due to greater uncertainty in the fitted value of R with decreasing $N_2$.

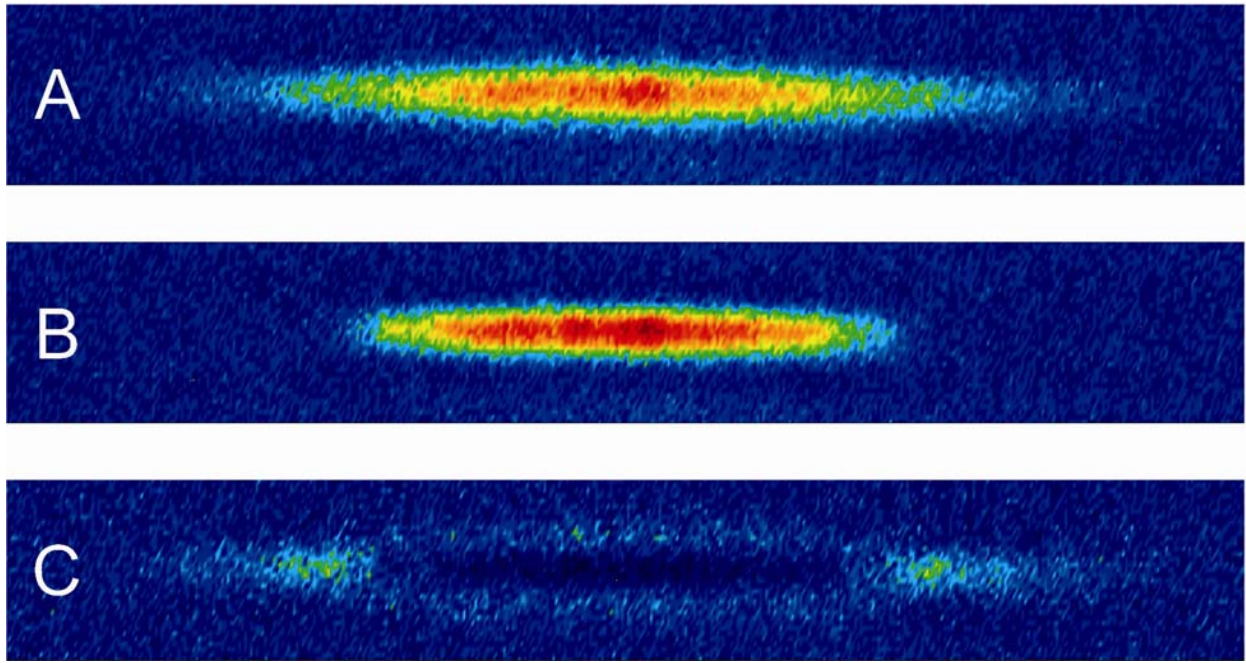

Figure 1

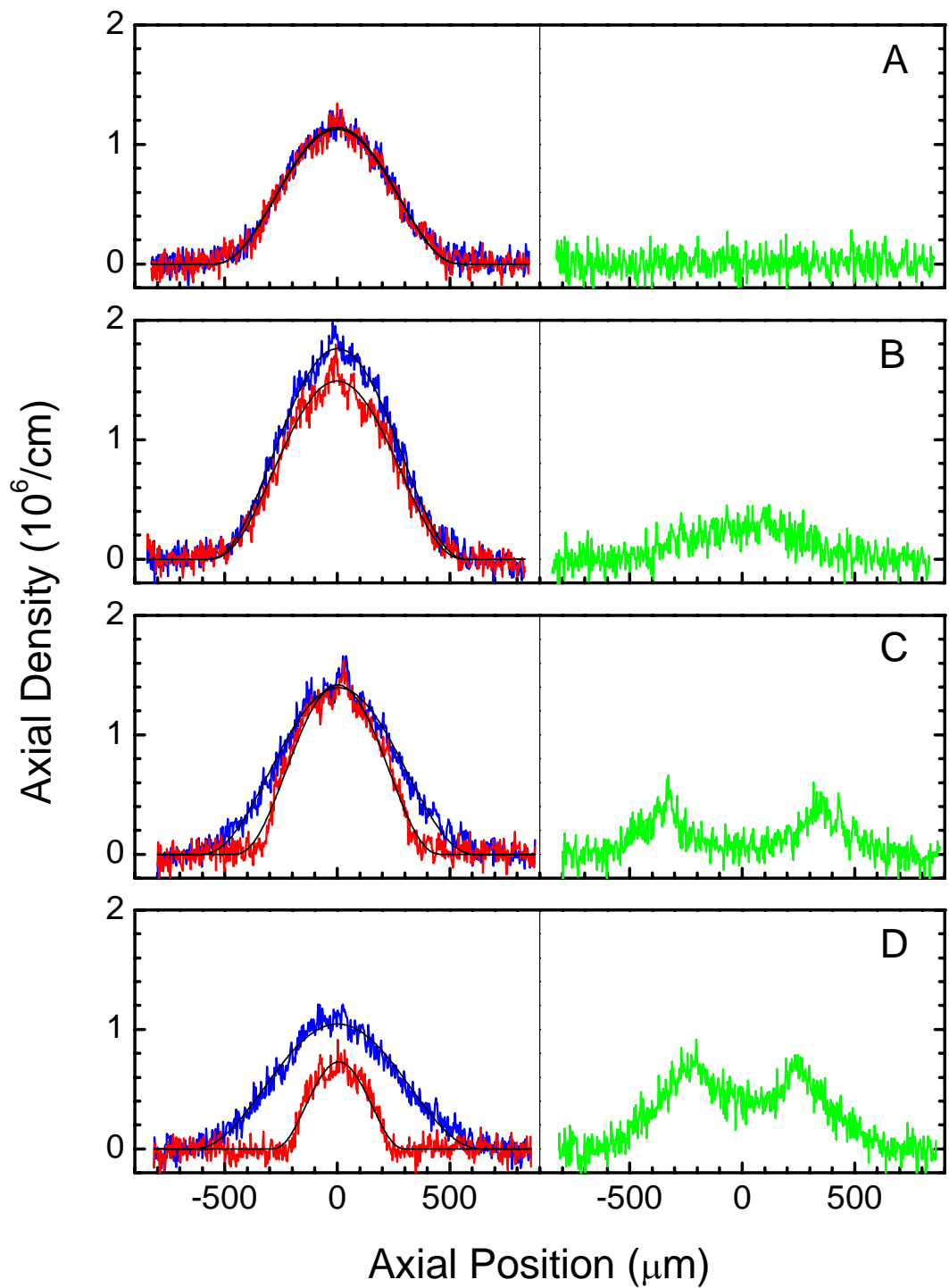

Figure 2

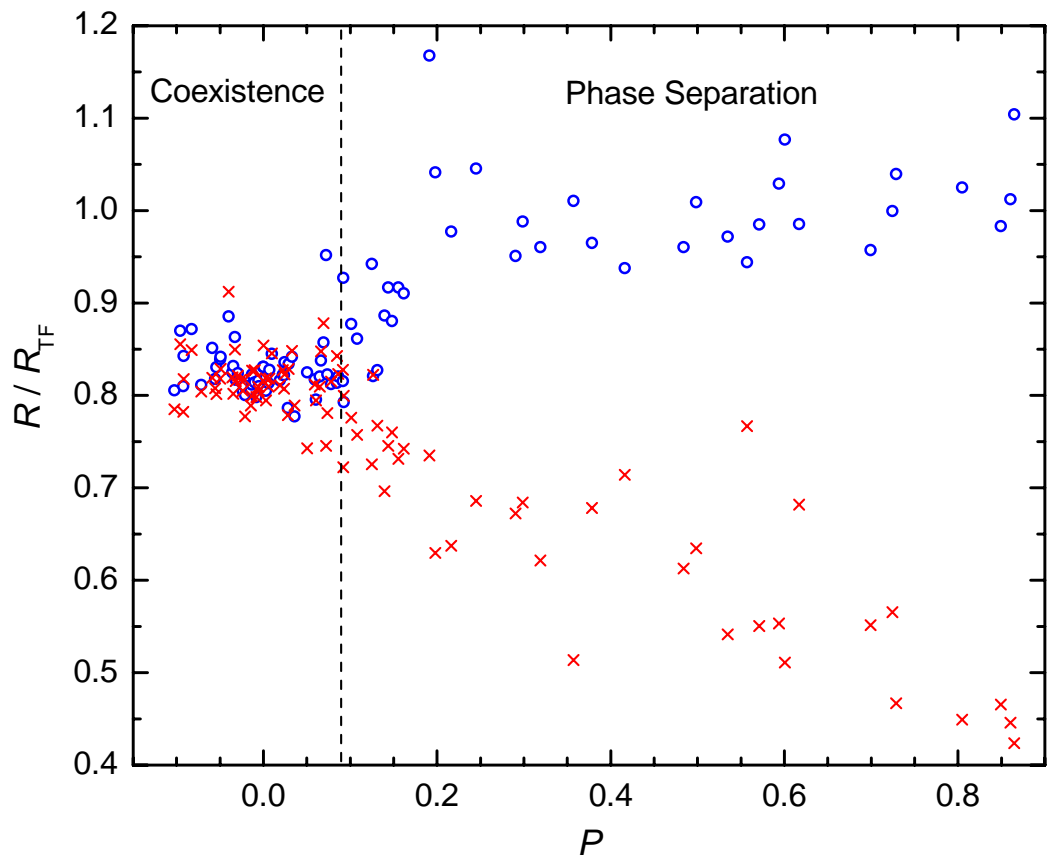

Figure 3

**Materials and methods:**

Our methods for producing a degenerate gas of fermionic $^6$Li atoms (*1, 2*) and the realization of the BEC-BCS crossover at a Feshbach resonance (*3*) have been described previously. Approximately $3 \times 10^6$ atoms at a temperature $T \approx 6$ μK are confined by an optical trap that is formed from a single focused infrared laser beam operating at a wavelength of 1080 nm. At an initial laser power of 2 W and with a $1/e^2$ beam radius of 26 μm, the trap depth is 90 μK and the corresponding radial and axial frequencies, $\upsilon_r$ and $\upsilon_z$, are 4.3 kHz and 40 Hz, respectively. The atoms are prepared in the energetically lowest Zeeman sublevel, $F = ½$, $m_F = ½$, in a nearly uniform bias field of 754 G. A series of 100 saw-tooth frequency sweeps, centered near 76 MHz, create an incoherent spin mixture of the $F = ½$, $m_F = ½$ (state $|1\rangle$) and the $F = ½$, $m_F = -½$ (state $|2\rangle$) sublevels. These states interact via a broad Feshbach resonance located near 834 G (*4, 5*). The relative numbers of atoms in $|1\rangle$ and $|2\rangle$ can be controlled by changing the power of the rf sweeps, thereby creating a polarized gas. By adjusting the power to transfer roughly 50% of the population in each sweep, a state with exactly $P = 0$ should be formed at the end of the 100 sweep sequence. Polarizations with $P > 0$ are achieved using less rf power.

After preparation of the spin mixture, the atoms are evaporatively cooled by reducing the optical trap depth over a period of 750 ms. Since the *s*-wave scattering length *a* is large near the Feshbach resonance, the elastic collision rate is high and evaporation is efficient. Evaporation continues until the trap depth reaches a final value of ~0.6 μK. The magnetic field is ramped to a desired field within the crossover region during the final 100 ms of evaporation. For the case of $P = 0$ at 754 G, a molecular Bose-Einstein condensate (MBEC) is observed to form with no detectable thermal molecules. A small magnetic field curvature produced by the magnetic bias coils contributes significantly to the axial confining potential for the low optical trap depths needed for cooling to the lowest temperatures (*3*). To accurately characterize the axial potential at each field, we directly measure $\upsilon_z$ by observing the period of oscillation of an atomic cloud that has been "kicked" by a transient field generated with an external coil. At 830 G, we find $\upsilon_r = 350$ Hz and $\upsilon_z = 7.2$ Hz at the final trap depth.

States $|1\rangle$ and $|2\rangle$ are sequentially and independently imaged in the trap by absorption using a probe laser beam resonant with the $^2S_{1/2}$ to $^2P_{3/2}$ atomic transition specific to each state. The two probes are each 5 μs in duration and are separated in time by 215 μs, which is fast compared to the timescale of oscillation in both the radial and axial dimensions, as well as to the expansion rate associated with the Fermi energy of the system. Although the probe frequencies are separated by 77 MHz, which is large compared to the transition linewidth of 5.9 MHz, a slight heating of the radial dimension is observed in the second image. This heating is due both to off-resonant excitation and to the release of binding energy from the dissociation of the weakly bound pairs. The second mechanism is only important in the BEC regime and does not contribute for fields above 800 G, where the molecular binding energy is less than 1 μK. At such fields, the axial profile obtained by integrating out the remaining radial direction (as shown in Fig. 2 of the report) shows no dependence on the probing order. However, for fields below 650 G, radial heating in the second image is so severe as to prevent detection of the second state.

Due to decreasing evaporation efficiency with increasing *P*, there is a correlation between *P* and total atom number. Figure S1 shows that $N_1$ decreases by a factor of two from $\sim 1.2 \times 10^5$ at $P = 0$ to $\sim 6 \times 10^4$ at $P = 0.2$, but is relatively constant for $P > 0.2$.

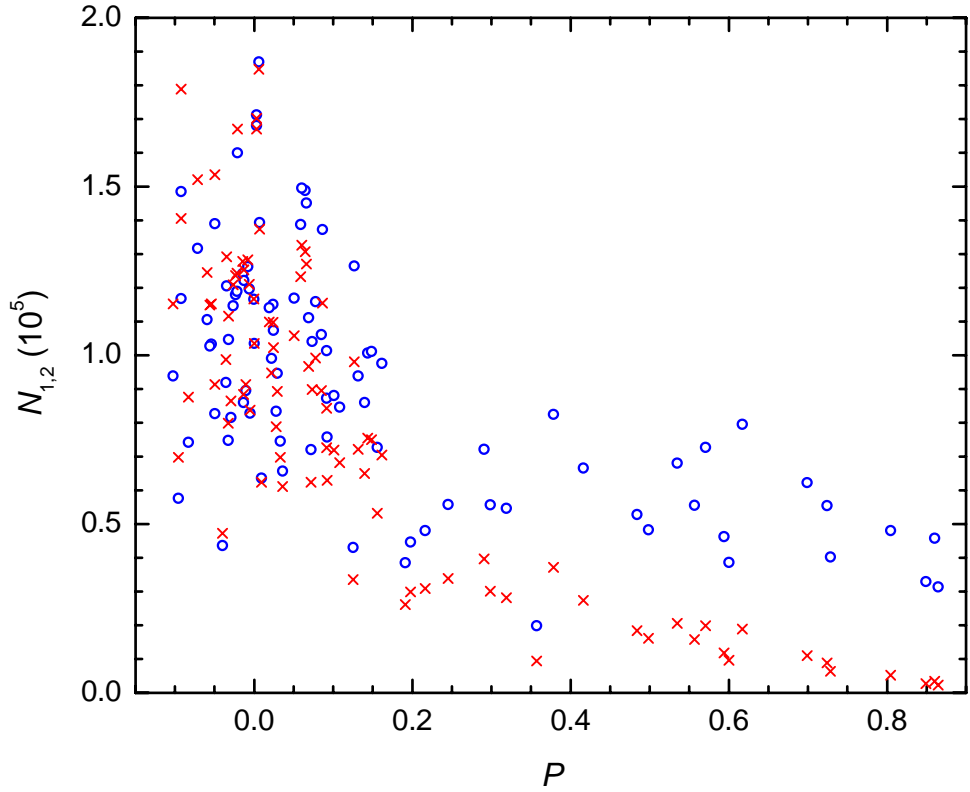

**Fig. S1.** $N_1$ and $N_2$ vs. $P$ at 830 G. Open circles correspond to $N_1$ while crosses correspond to $N_2$. The correlation between $N_1$, $N_2$, and $P$, is attributed to decreasing evaporation efficiency in the optical trap for increasing polarization. For larger $P$ values, smaller numbers result in increased relative uncertainties in $P$.

1. A. G. Truscott, K. E. Strecker, W. I. McAlexander, G. B. Partridge, R. G. Hulet, *Science* **291**, 2570 (2001).
2. K. E. Strecker, G. B. Partridge, R. G. Hulet, *Phys. Rev. Lett.* **91**, 080406 (2003).
3. G. B. Partridge, K. E. Strecker, R. I. Kamar, M. W. Jack, R. G. Hulet, *Phys. Rev. Lett.* **95**, 020404 (2005).
4. M. Houbiers, H. T. C. Stoof, W. I. McAlexander, R. G. Hulet, *Phys. Rev. A* **57**, R1497 (1998).
5. M. Bartenstein *et al.*, *Phys. Rev. Lett.* **94**, 103201 (2005).